\documentclass[prd,preprint,tightenlines,floatfix,showpacs,preprintnumbers,nofootinbib,eqsecnum]{revtex4}

 \usepackage[dvips,final]{graphicx}
  \usepackage{amssymb}
   \usepackage{amsmath}
    \usepackage{amsfonts}
     \usepackage{epsfig}
      \usepackage{bm} % bold math

\begin{document}

%\begin{flushright}
%\vskip1cm
%\end{flushright}

\title{Anomalous current from covariant Wigner functions}

\author{George Prokhorov$^1$}
\email{prokhorov@theor.jinr.ru}

\author{Oleg Teryaev$^1$}
\email{teryaev@theor.jinr.ru}

\affiliation{
{$^1$\sl 
%Bogoliubov Laboratory of Theoretical Physics, 
Joint Institute for Nuclear Research, Dubna, Russia
}\vspace{1.5cm}
}
%
%
%
%%%%%%%%%%%%%%%%%%%%%%%%%%%%%%%%%%%%%%%%%%%%%%%%%%%%
\begin{abstract}
\vspace{0.5cm}
We consider accelerated and rotating media of weakly interacting fermions in local thermodynamic equilibrium. Kinetic properties of this media are described by covariant Wigner function calculated on the basis of relativistic distribution function of particles with spin. We obtain the formulae for axial current beyond the approximation of smallness of thermal vorticity tensor and chemical potential and calculate its divergence. In the massless limit higher order terms in vorticity and chemical potential compensate each other  with only the low-order contributions surviving. It is shown, that axial current gets a topological component along the 4-acceleration vector. The similarity between different approaches to baryon polarisation is established. 
\end{abstract}
%%%%%%%%%%%%%%%%%%%%%%%%%%%%%%%%%%%%%%%%%%%%%%%%%%%%

%\pacs{}

\maketitle

%================
%\section{Notations}
%================

%================
\section{Introduction}
%================

Relativistic fermionic liquid is an unusual object for which the laws of quantum field theory become apparent in macroscopic phenomena. Thus two important effects from this series, chiral magnetic effect (CME) and chiral vortical effect (CVE), which are essentially quantum field effects \cite{Kharzeev:2012ph}(volume of Springer lectures in physics), \cite{Son:2009tf}, \cite{Sadofyev:2010is}, \cite{Vilenkin:1979ui}, \cite{Vilenkin:1980fu}, \cite{Rogachevsky:2010ys}, \cite{Zakharov:2012vv},  effectively modify equations of hydrodynamics \cite{Son:2009tf}. Chiral effects can have different experimental consequences, in particular resulting in baryon polarisation \cite{Rogachevsky:2010ys, Baznat:2017ars}, approached also in seemingly different context \cite{Becattini:2016gvu}, \cite{Becattini:2013fla}. In particular, there are suggestions to investigate P-odd effects in heavy-ion collisions, strongly controlled by CVE \cite{Rogachevsky:2010ys,Baznat:2017ars}, at the Nuclotron-Based Ion Collider Facility (NICA) at the Joint Institute for Nuclear Research. 

There are currently seemingly different approaches to polarisation relying either on relativistic thermodynamics \cite{Becattini:2013fla}
or on anomalous axial current and the respective charge \cite{Rogachevsky:2010ys,Baznat:2017ars}. The use of that charge allows to 
address the baryonic phase performing the calculation of quark anomalies. The other approach to polarisation in confined phase was recently suggested by consideration of the vortices cores in pionic superfluid \cite{Teryaev:2017nro}. To compare the approaches and fill the gap between them we explore in detail the appearance of anomalous axial current (used  in \cite{Rogachevsky:2010ys,Baznat:2017ars}) in the framework of the approach \cite{Becattini:2013fla}. 

To do so we will concentrate on the (axial) CVE. The existence of CVE was proven in hydrodynamic approximation as the result of joint consideration of triangle axial anomaly and second law of thermodynamics \cite{Son:2009tf}. It also can be obtained from the calculation of triangle loop diagrams with external topological field \cite{Sadofyev:2010is}, \cite{Rogachevsky:2010ys}. 

We will consider the media of fermions in approximation of local thermodynamic equilibrium. The basis of our analysis was laid in the series of papers \cite{Becattini:2013fla},\cite{Becattini:2009wh},\cite{Becattini:2016gvu},\cite{Becattini:2011ev},\cite{Becattini:2007nx},\cite{Becattini:2007zn},\cite{Buzzegoli:2017cqy}, in which relativistic local distribution functions for massive particles with spin were introduced. In \cite{Becattini:2013fla} it was shown, that these distribution functions lead to the corrections on the vorticity to the thermodynamic quantities, and the terms of the first orders over the vorticity were derived for energy-momentum tensor, vector current, spin tensor and recently in \cite{Buzzegoli:2017cqy} for axial current. In this paper we continue to consider the consequences of these distribution functions and define accurate analytic formulae for axial beyond the approximation of the vorticity smallness.

At first, following to \cite{Buzzegoli:2017cqy}, we show that CVE obtained from quantum effective field theory \cite{Sadofyev:2010is}, also follows from the consideration of relativistic distribution functions (notice, that this result was also obtained in \cite{Gao:2012ix},\cite{Gao:2017gfq} for modified Wigner function), which already implies the similarity of two approaches. 
This may be explained by implicit apperance of triangle graph \footnote{We are indebted for V.I. Zakharov for pointing out of such an effect}. 
On the basis of the obtained accurate formula for axial current we show, that in massless limit higher order effects of the vorticity and chemical potential are compensated, leaving only the terms up to the third order. An intriguing consequence of the obtained formulae, is that axial current obtains a topological component along 4-acceleration vector, which gives a contribution to axial current divergence.

%The comparison of the obtained results with previous calculations is carried out.

We use Minkowskian metric tensor in the form $g_{\mu\nu}=\mathrm{diag}(1,-1,-1,-1)$ and Levi-Civita symbol $\epsilon^{0123}=1$. The contraction of the induces is sometimes denoted by dots $\varpi^2=\varpi_{\mu\nu}\varpi^{\mu\nu}=\varpi:\varpi$. For Dirac matrices we take $\gamma^5=i \gamma^0\gamma^1\gamma^2\gamma^3$. We use the system of units $\hbar=c=k=1$.
%=================================================
\section{Thermal vorticity tensor}
\label{Sec:Thermal vorticity tensor}
%=================================================

Thermal vorticity tensor \cite{Becattini:2013fla}, \cite{Buzzegoli:2017cqy} contains information about local acceleration, vorticity and temperature gradients in the media with local thermodynamic equilibrium. As it was noticed in \cite{Becattini:2013fla} the form of this tensor is not strictly defined for local thermodynamic equilibrium and known up to the second order over the gradients $\partial^2\beta$
\begin{eqnarray}
\varpi_{\mu\nu}= -\frac{1}{2}(\partial_{\mu}\beta_{\nu}-\partial_{\nu}\beta_{\mu})+\mathcal{O}(\partial^2\beta)\,,
\label{thermal vort tensor}
\end{eqnarray}
where $\beta_{\mu}=\frac{u_{\mu}}{T}$ and $T$ is a temperature in the comoving system. In the limit of global thermodynamic equilibrium $\beta_{\mu}$ satisfies \cite{Buzzegoli:2017cqy}
\begin{eqnarray}
\beta_{\mu}= b_{\mu}+\varpi_{\mu\nu}x_{\nu}\,,\quad b_{\mu}=\mathrm{const}\,,\quad\, \varpi_{\mu\nu}=\mathrm{const}\,,\quad \varpi_{\mu\nu}= -\frac{1}{2}(\partial_{\mu}\beta_{\nu}-\partial_{\nu}\beta_{\mu})\,
\label{b global}
\end{eqnarray}
and relation of chemical potential to temperature in comoving frame is constant
\begin{eqnarray}
\xi=\frac{\mu}{T}=\mathrm{const}\,.
\label{xi}
\end{eqnarray}
Fulfilment of (\ref{b global}), (\ref{xi}) transforms density operator to the form of global thermodynamic equilibrium density operator \cite{Buzzegoli:2017cqy}. In most of the calculations we will not need the exact accurate form of $\varpi$. In fact it is sufficient  to suppose that thermal vorticity is an antisymmetric tensor, as it follows already from the form of the distribution functions. But at the final stages, we will consider global thermodynamic equilibrium (\ref{b global}) as a particular case to illustrate the content and the consequences of the obtained formulae.

Similarly to electrodynamics it is convenient to decompose $\varpi$ tensor to vector and pseudovector. Following to the paper \cite{Buzzegoli:2017cqy} we introduce thermal acceleration vector $\alpha_{\mu}$ and thermal vorticity pseudovector $w_{\mu}$
\begin{eqnarray}
\alpha_{\mu}= \varpi_{\mu\nu} u^{\nu},\quad w_{\mu}= -\frac{1}{2}\epsilon_{\mu\nu\alpha\beta}u^{\nu}\varpi^{\alpha\beta}\,.
\label{alphaw}
\end{eqnarray}
Linking to electrodynamics, $\alpha_{\mu}$ vector corresponds to "electric" component in comoving frame and $w_{\mu}$ corresponds to comoving "magnetic" component. Then tensor $\varpi_{\mu\nu}$ can be decomposed to (\ref{alphaw})
\begin{eqnarray}
\varpi_{\mu\nu}= \epsilon_{\mu\nu\alpha\beta}w^{\alpha}u^{\beta}+\alpha_{\mu}u_{\nu}-\alpha_{\nu}u_{\mu}\,.
\label{decomp}
\end{eqnarray}
For global thermodynamic equilibrium temperature remains constant along flow direction \cite{Buzzegoli:2017cqy}. This allows to prove, that $\alpha_{\mu}$ and $w_{\mu}$ will be proportional to 4-acceleration and vorticity in this case
\begin{eqnarray}
\alpha_{\mu}= \frac{1}{T}u^{\nu}\partial_{\nu}u_{\mu}=\frac{a_{\mu}}{T},\quad w_{\mu}= \frac{1}{2T}\epsilon_{\mu\nu\alpha\beta}u^{\nu}\partial^{\alpha}u^{\beta}=\frac{\omega_{\mu}}{T}\,.
\label{alphaw global}
\end{eqnarray}
In the local rest frame $a$ and $\omega$ are expressed by ordinary three-dimensional acceleration and rotation speed
\begin{eqnarray}
a^{\mu}= (0,\vec{a}),\quad \omega^{\mu}= (0,\vec{\omega})\,,
\label{alphaw global rest}
\end{eqnarray}
where $\vec{a}$ and $\vec{\omega}$ are 3-dimensional acceleration and angular velocity. In the limit (\ref{b global}),(\ref{xi}) the divergences and derivative of temperature will be \cite{Buzzegoli:2017cqy}
\begin{eqnarray}
\partial\cdot \alpha = \frac{1}{|\beta|}(2w^2-\alpha^2),\quad w_{\mu}= -\frac{3}{|\beta|}(w\cdot \alpha),\quad \partial_{\mu}T=T^2\alpha_{\mu}\,.
\label{div a w dT}
\end{eqnarray}
It is possible to construct tensor $\widetilde{\varpi}$, which is dual to $\varpi$ and gives vorticity vector after projection to 4-velocity 
\begin{eqnarray}
\widetilde{\varpi}_{\mu\nu}=\frac{1}{2}\epsilon_{\mu\nu\alpha\beta}\varpi^{\alpha\beta}= \epsilon_{\mu\nu\alpha\beta}\alpha^{\alpha}u^{\beta}-w_{\mu}u_{\nu}+w_{\nu}u_{\mu}\,,\quad
w_{\mu}=\widetilde{\varpi}_{\nu\mu}u^{\nu}\,.
\label{dual decomp}
\end{eqnarray}
Analogically to electrodynamics one scalar and one pseudoscalar can be constructed from the thermal vorticity tensor (\ref{thermal vort tensor}) $\varpi^2=\varpi:\varpi$ and $\varpi :\widetilde{\varpi}$, respectively
\begin{eqnarray}
\varpi^2=2(\alpha^2-w^2),\quad \varpi :\widetilde{\varpi}=-4(w\cdot\alpha)
\label{w2wwv}
\end{eqnarray}
Notice, that $\varpi^2=\mathrm{const}$, $\varpi :\widetilde{\varpi}=\mathrm{const}$ in the limit of global thermodynamic equilibrium as it follows from (\ref{b global}). Another scalar, which will appear,  corresponds to Hamiltonian density of electromagnetic fields
\begin{eqnarray}
\alpha^2+w^2\,.
\label{a2plusw2}
\end{eqnarray}
Similarly to electrodynamics it is convenient to introduce complex vectors for (\ref{alphaw global})
\begin{eqnarray}
\varphi_{\mu}=\frac{a_{\mu}}{2\pi}+\frac{i\omega_{\mu}}{2\pi}\,,\quad \psi_{\mu}=\varphi_{\mu}^*\,.
\label{complex}
\end{eqnarray}

%=================================================
\section{Covariant Wigner function and distribution function for particles with spin}
\label{Sec:Wigner}
%=================================================

One of the methods to describe kinetic properties of the media, which allows one to take into account quantum effects, is based on covariant Wigner function. For the spin 1/2 particles this is a spinorial matrix, expressed by mean value of the operators of Dirac fields
\begin{eqnarray}
W(x,k)_{AB}= -\frac{1}{(2\pi)^4}\int d^4y e^{-ik\cdot y}\langle :\Psi_{A}(x-y/2)\overline {\Psi}_{B}(x+y/2):\rangle\,.
\label{Wpsipsi}
\end{eqnarray}
Brackets $\langle::\rangle$ mean ensemble averaging with normal ordering. Wigner function (\ref{Wpsipsi}) depends on distribution function \cite{Becattini:2013fla}
\begin{eqnarray}W(x,k)= \frac{1}{2}\int \frac{d^3 p}{\varepsilon} \Big(\delta^{4}(k-p)U(p)f(x,p)\bar{U}(p)- \delta^{4}(k+p)V(p)\bar{f}^{\,T}(x,p)\bar{V}(p)\Big)
\label{Wf}
\end{eqnarray}
where $U(p)=(u_{+}(p),u_{-}(p))$, $V(p)=(v_{+}(p),v_{-}(p))$ are $4\times 2$ matrices, $\bar{U}(p)=U^{\dag}(p)\gamma^{0}$, $\bar{V}(p)=V^{\dag}(p)\gamma^{0}$ are $2\times 4$ matrices, $f(x,p)$ and $\bar{f}(x,p)$ are $2\times 2$ matrices and $u_{+}(p)$ and $u_{-}(p)$ are spinors of free Dirac fields with different values of helicity, normalised as usually $\bar{u}_r u_{s}=-\bar{v}_rv_{s}=2m\delta_{rs}$.

In \cite{Becattini:2013fla} local equilibrium distribution functions $f(x,p)$ and $\bar{f}(x,p)$ for massive particles with spin in the accelerated media were introduced. They have the following form
\begin{eqnarray}
f(x,p)=\frac{1}{8\pi^3}\frac{1}{2m}\bar{U}(p)X(x,p)U(p)\,,\quad 
\bar{f}(x,p)=-\frac{1}{8\pi^3}\frac{1}{2m}\big[\bar{V}(p)\bar{X}(x,p)V(p)\big]^{T}\,
\label{fX}
\end{eqnarray}
(note the phase space factor $\frac{1}{8\pi^3}$, c.f. \cite{Cooper:1974mv} and \cite{Buzzegoli:2017cqy}). Functions $X(x,p)$ and $\bar{X}(x,p)$ have the form
\begin{eqnarray}\nonumber
&& X(x,p)=\Big(\exp[\beta\cdot p-\xi(x)]\exp\Big[-\frac{1}{2}\varpi(x) :\Sigma\Big]+I\Big)^{-1}\,, \\
&& \bar{X}(x,p)=\Big(\exp[\beta\cdot p+\xi(x)]\exp\Big[\frac{1}{2}\varpi(x) :\Sigma\Big]+I\Big)^{-1}\,
\label{X}
\end{eqnarray}
In (\ref{X}) $\xi(x)=\frac{\mu(x)}{T(x)}$, $\mu(x)$ and $T(x)$ are comoving local chemical potential and temperature and $\Sigma_{\mu\nu}=\frac{i}{4}[\gamma_{\mu},\gamma_{\nu}]$ are the generators of Lorentz transformation of spinors.

Mean values can be now calculated with a help of Wigner functions $W(x,p)$, and we will now concentrate on the axial current.
%=================================================
\section{Axial current}
\label{Sec:axial}
%=================================================

%=================================================
\subsection{Massive particles}
%=================================================

In general, mean value of the current $\langle:\bar{\Psi}A\Psi\rangle$, where $A$ is an operator, containing $4\times 4$ matrix, has the form \cite{Becattini:2013fla}
\begin{eqnarray} \nonumber
&& \langle:\bar{\Psi}(x)A\Psi(x):\rangle=\int d^4k\, \mathrm{tr}(AW(x,k))= \\
&& \int \frac{d^3p}{2\varepsilon}\mathrm{tr}_2\big(f(x,p)\bar{U}(p)AU(p)\big)-\mathrm{tr}_2\big(\bar{f}^{\,T}(x,p)\bar{V}(p)AV(p)\big)\,
\label{meanv currents}
\end{eqnarray}
where $\mathrm{tr}_2$ means the trace of $2\times 2$ matrix, corresponding to sum over polarisations. For axial current $j_{\mu}^{5}=\bar{\Psi}\gamma_{\mu}\gamma^{5}\Psi$ (\ref{meanv currents}) leads to
\begin{eqnarray}\langle :j_{\mu}^{5}:\rangle = -\frac{1}{16\pi^3} \epsilon_{\mu\alpha\nu\beta} \int \frac{d^3p}{\varepsilon} p^{\alpha}\Big\{
\mathrm{tr}\big(X\Sigma^{\nu\beta}\big) -
\mathrm{tr}\big(\bar{X}\Sigma^{\nu\beta}\big)\Big\}\,.
\label{axial mean}
\end{eqnarray}
The calculation of the traces in (\ref{axial mean}) can be made accurately as it is shown in Appendix \ref{Sec:trace}. The integral in (\ref{axial mean}) has a form $\int \frac{d^3 p}{\varepsilon}p^{\alpha} f(\beta\cdot p)$, where $f(\beta\cdot p)$ is a scalar function of $(\beta\cdot p)$. From Lorenz-covariance one obtains
\begin{eqnarray}
\int \frac{d^3 p}{\varepsilon}p^{\alpha} f(\beta\cdot p)=u^{\alpha}\int d^3 p f(\frac{\varepsilon}{T})\,.
\label{b int}
\end{eqnarray}
After substitution of (\ref{trXSigma3}) to (\ref{axial mean}), using (\ref{b int}) and after algebraic transformations, one can see that imaginary terms compensate each other and we obtain final expression for axial current 
\begin{eqnarray} 
&& \langle :j_{\mu}^{5}:\rangle = C_1 w_{\mu} +\mathrm{sgn}(\varpi:\widetilde{\varpi})\, C_2 \alpha_{\mu}\,,
\label{axial current main} \\ \nonumber
&& C_1=\frac{1}{4 \pi^2}\frac{g_2 \cosh(g_1)\sin(g_2)+g_1\sinh(g_1)\cos(g_2)}{g_1^2+g_2^2}\big(I_1(\xi)+I_1(-\xi)\big)+ \\ \nonumber
&& \frac{1}{8 \pi^2}\frac{g_1 \sinh(2g_1)+g_2\sin(2g_2)}{g_1^2+g_2^2}\big(I_2(\xi)+I_2(-\xi)\big)\,, \\ \nonumber
&& C_2=\frac{1}{4 \pi^2}\frac{g_2 \sinh(g_1)\cos(g_2)-g_1\cosh(g_1)\sin(g_2)}{g_1^2+g_2^2}\big(I_1(\xi)+I_1(-\xi)\big)+ \\ \nonumber
&& \frac{1}{8 \pi^2}\frac{g_2 \sinh(2g_1)-g_1\sin(2g_2)}{g_1^2+g_2^2}\big(I_2(\xi)+I_2(-\xi)\big)\,,
\end{eqnarray}
where $g_1$ and $g_2$ are scalars, depending on vorticity and given by (\ref{sqrt}). $I_1(\xi)$, $I_2(\xi)$ denote one-dimensional integrals, which can be calculated numerically:
\begin{eqnarray} \nonumber
&& I_1(\xi)=\int\frac{dp\, \vec{p}^{\;2}\cosh(\frac{\varepsilon}{T}-\xi)}{ \Big(\cosh(\frac{\varepsilon}{T}-\xi-g_1)+\cos(g_2)\Big)\Big(\cosh(\frac{\varepsilon}{T}-\xi+g_1)+\cos(g_2)\Big)}\,,\\ 
&& I_2(\xi)=\int\frac{dp\, \vec{p}^{\;2}}{ \Big(\cosh(\frac{\varepsilon}{T}-\xi-g_1)+\cos(g_2)\Big)\Big(\cosh(\frac{\varepsilon}{T}-\xi+g_1)+\cos(g_2)\Big)}\,.
\label{axial current main int}
\end{eqnarray}

The expression (\ref{axial current main}) contains two different terms. The first one with $w_{\mu}$ corresponds to usual expression for CVE \cite{Son:2009tf}, \cite{Zakharov:2012vv}, \cite{Buzzegoli:2017cqy}. At the same time, in comparison to the usual formula for CVE, (\ref{axial current main int}) also includes "electric" term, proportional to 4-vector $\alpha_{\mu}$, corresponding to acceleration vector, divided to temperature in the case of global thermodynamic equilibrium. As $C_1=O(1)$, while $C_2=O(g_1 g_2)=O(\varpi:\widetilde{\varpi})$, the term with  $\alpha_{\mu}$ is of the third order over the vorticity.

%=================================================
\subsection{Massless limit}
%=================================================

Massless limit is of special interest, as it corresponds to chiral invariance manifestation, and also because in this case the expressions (\ref{axial current main}), (\ref{axial current main int}) can be significantly simplified and integrals (\ref{axial current main int}) can be calculated analytically. For this purpose it is convenient to come back to the formulae (\ref{trXSigma3}). Using (\ref{b int}), (\ref{axial mean}), (\ref{trXSigma3}) and coming to the massless limit $\varepsilon=|\vec{p}|$, the integrals in (\ref{axial mean}) can be expressed through polylogarithmic functions $\mathrm{Li}_3(z)$
\begin{eqnarray} \nonumber
&& \langle :j_{\mu}^{5}:\rangle =-\frac{T^3}{4\pi^2 (g_1^2+g_2^2)}\Big((g_1+ig_2)\mathrm{Li}_3(-e^{g_1-ig_2-\xi})-(g_1-i g_2)\mathrm{Li}_3(-e^{-g_1-ig_2-\xi})+\\ \nonumber
&& (g_1+ig_2)\mathrm{Li}_3(-e^{g_1-ig_2+\xi})-(g_1-i g_2)\mathrm{Li}_3(-e^{-g_1-ig_2+\xi})+ c.c.\Big)w_{\mu}+\\ 
&& \frac{\mathrm{sgn}(\varpi:\widetilde{\varpi})T^3}{4\pi^2 (g_1^2+g_2^2)}\Big((g_2+ig_1)\mathrm{Li}_3(-e^{-g_1-ig_2-\xi})-(g_2-i g_1)\mathrm{Li}_3(-e^{g_1-ig_2-\xi})+\label{axial current polylog} \\ \nonumber
&& (g_2+ig_1)\mathrm{Li}_3(-e^{-g_1-ig_2+\xi})-(g_2-i g_1)\mathrm{Li}_3(-e^{g_1-ig_2+\xi})+c.c.\Big)\alpha_{\mu}\,.
\end{eqnarray}
The polylogarithms $\mathrm{Li}_3(z)$ in (\ref{axial current polylog}) have the notable property \cite{Prudnikov}
\begin{eqnarray}
\mathrm{Li}_3(-e^{-x})-\mathrm{Li}_3(-e^{x})=\frac{\pi^2}{6}x+\frac{1}{6}x^3\,,
\label{Li}
\end{eqnarray}
which allows one to simplify (\ref{axial current polylog}) significantly. Using the definitions of $g_1$ and $g_2$ (\ref{sqrt}) and formulae (\ref{w2wwv}) one obtains
\begin{eqnarray}
\langle :j_{\mu}^{5}:\rangle =\big(\frac{T^2}{6}[1+\frac{\alpha^{2} -w^2}{4\pi^2}]+\frac{\mu^2}{2\pi^2}\big)Tw_{\mu}+\frac{T^3}{12\pi^2}(w\cdot\alpha)\,\alpha_{\mu}\,.
\label{axial current awT}
\end{eqnarray}
In the limit of global thermodynamic equilibrium, using (\ref{alphaw global}), (\ref{axial current awT}) can be simplified
\begin{eqnarray}
\langle :j_{\mu}^{5}:\rangle =\Big(\frac{1}{6}\big[T^2+\frac{a^{2} -\omega^2}{4\pi^2}\big]+\frac{\mu^2}{2\pi^2}\Big)\omega_{\mu}+\frac{1}{12\pi^2}(\omega\cdot a)\,a_{\mu}\,.
\label{axial current aw}
\end{eqnarray}
The divergence of (\ref{axial current awT}) in the case of global thermodynamic equilibrium can be defined in straightforward way, using the formulae from \cite{Buzzegoli:2017cqy}, in particular (\ref{div a w dT})
\begin{eqnarray}
&& \partial^{\mu}\langle :j_{\mu}^{5}:\rangle=\frac{1}{6\pi^2}(\omega\cdot a)(a^2 +\omega^2) \,.
\label{axial divergence}
\end{eqnarray}
The contribution to the right hand side of (\ref{axial divergence}) comes from the last acceleration term in (\ref{axial current aw}). This contribution is purely topological being proportional to $(\omega\cdot a)$, it is of the forth order over the vorticity and it vanishes in the limit $a\to 0$.

Formula (\ref{axial current aw}) can be separated to three parts: thermal vortical current $\langle :j_{\mu}^{5}:\rangle_{\mathrm{Tvort}} $, depending on temperature, chemical potential and vorticity, vortical term $\langle :j_{\mu}^{5}:\rangle_{\mathrm{vort}}$, which doesn't depend on $\mu$ and $T$, and topological term $\langle :j_{\mu}^{5}:\rangle_{\mathrm{top}} $ , expressed only through vorticity and acceleration
\begin{eqnarray} \nonumber
&& \langle :j_{\mu}^{5}:\rangle =\langle :j_{\mu}^{5}:\rangle_{\mathrm{Tvort}}+\langle :j_{\mu}^{5}:\rangle_{\mathrm{vort}} +\langle :j_{\mu}^{5}:\rangle_{\mathrm{top}}\,,\\ \nonumber
&& \langle :j_{\mu}^{5}:\rangle_{\mathrm{Tvort}} =\Big(\frac{T^2}{6}+\frac{\mu^2}{2\pi^2}\Big)\omega_{\mu}\,,\quad
\langle :j_{\mu}^{5}:\rangle_{\mathrm{vort}}= \frac{a^{2} -\omega^2}{24\pi^2}\omega_{\mu}\,, \nonumber \\
&& \langle :j_{\mu}^{5}:\rangle_{\mathrm{top}} = \frac{1}{12\pi^2}(\omega\cdot a)\,a_{\mu}\,.
\label{axial current therm top}
\end{eqnarray}
In the limit $T\,,\,\mu\to 0$ the first term $\langle :j_{\mu}^{5}:\rangle_{\mathrm{Tvort}}$ vanishes, while the last two terms in (\ref{axial current therm top}) form nonzero contribution. Such a structure of a current maight  be related to the vacuum effects  in accelerated frames \cite{Zakharov:2015taa}.

Let's notice, that introduction of complex vectors (\ref{a2plusw2}) allows one to diagonalize expression (\ref{axial current aw})
\begin{eqnarray}
\langle :j_{\mu}^{5}:\rangle =2\pi\, \mathrm{Im}\Big[\big(\frac{1}{6}(T^2+\varphi^2)+\frac{\mu^2}{2\pi^2}\big)\varphi_{\mu}\Big]\,.
\label{axial current diag}
\end{eqnarray}
Expression (\ref{axial current aw}) for CVE in linear approximation over the vorticity and acceleration corresponds to the results from effective field theory calculations \cite{Sadofyev:2010is}, \cite{Rogachevsky:2010ys}. Hence the coincidence of effective field theory and kinetic approaches is established (this fact was also shown in \cite{Buzzegoli:2017cqy},\cite{Gao:2012ix}).

Equations (\ref{axial current awT}),(\ref{axial current aw}) were obtained as the result of accurate integration of distribution functions in (\ref{axial mean}) without expansion over the vorticity. But at the end we have obtained formulae (\ref{axial current awT}),(\ref{axial current aw}) with terms of only up to the third order over the vorticity and chemical potential. This means that higher order terms are compensated by each other. The origin of this fact can be seen from the formulae (\ref{Li}). The contributions of particles and antiparticles in (\ref{axial current polylog}) correspond to polylogarithms with opposite signs in the exponent factor: for example, there is the term of the form $-(g_1+i g_2)\mathrm{Li}_3(-e^{-g_1+ig_2+\xi})$ and $(g_1+ig_2)\mathrm{Li}_3(-e^{g_1-ig_2-\xi})$. Both of these two terms can be expanded to infinite Taylor series over the vorticity and chemical potential. But from (\ref{Li}) one can see, that these two terms compensate each other starting from the fourth order. So the correspondence of (\ref{axial current awT}),(\ref{axial current aw}) to the low order expansion over the vorticity and chemical potential is the result of compensation between different contributions from the traces in (\ref{axial current polylog}).

The expression (\ref{axial current aw}) can be compared to the existing results. In particular, in \cite{Vilenkin:1979ui} axial current of right handed massless fermions was calculated for the system with constant rotation speed on the axis of rotation. In the limit $a_{\mu}=0$ (\ref{axial current aw}) exactly corresponds to Eq.(27) from \cite{Vilenkin:1979ui}, where the term independent of $T,\mu$ was first identified. But in comparison to \cite{Vilenkin:1979ui} we find the term along $a_{\mu}$, which in \cite{Vilenkin:1979ui} equals zero, because rotation speed is perpendicular to centripetal acceleration. Also we find the component along $\omega_{\mu}$, which is quadratic over acceleration. In \cite{Buzzegoli:2017cqy} axial current was calculated up to the first order over the vorticity which is also in accordance with (\ref{axial current aw}). 
%The formula (\ref{axial current aw}) also coincides to pioneering paper \cite{Son:2009tf}.

%=================
\section{Conclusion}
\label{Sec:Conclusion}
%=================

In this paper we have considered the results, which can be obtained from the Wigner and distribution functions introduced in \cite{Becattini:2013fla}. We have obtained analytic formulae for axial current both for massive (\ref{axial current main}) and massless particles (\ref{axial current awT}), (\ref{axial current aw}), proving the coincidence with effective field theory approach. For massless fermions it was possible to exceedingly simplify the expressions due to compensation of higher order terms over the vorticity and chemical potential. It is shown, that axial current remains nonzero in the limit $T\,,\,\mu\to 0$ (\ref{axial current therm top}). In particular, it contains
the new topological component along the vector $\alpha_{\mu}$. 
In the limit of global thermodynamic equilibrium this means that the current will be directed along the acceleration vector.  
Also we calculate the divergence of average axial current and suggest simple way of diagonalization of axial current through introduction of complex superposition of acceleration and vorticity.

Finally, we may conclude that various approaches to baryon polarizations are closely related and that found earlier numarical similarity between them \cite{Baznat:2017ars} may be not an occasional one. 

{\bf Acknowledgments}

Useful discussions with A.S. Sorin and V.I. Zakharov are gratefully acknowledged.
The work was supported in part by Russian Science Foundation
Grant No 16-12-10059.

%===========================

\appendix

%=============================================
\section{Trace calculation}
\label{Sec:trace}
%=============================================

The present goal is to obtain exact formulae for the traces in (\ref{axial mean}). For this purpose it is necessary to expand $X(x,p)$ to Taylor series, take the trace in an every term and then sum the trace back. Function $\frac{1}{1+x}$ can be expanded to Taylor series
\begin{eqnarray}
\frac{1}{1+x}=\sum^{\infty}_{n=0}(-1)^n x^{t(n+l)}\,,\quad t,l\Rightarrow
\left\{ \begin{array}{ll}
t=1,l=0 & \mathrm{if}\, |x|<1\,,\\
t=-1,l=1 & \mathrm{if}\, |x|>1\,.
\end{array} \right.
\label{taylor}
\end{eqnarray}
Then for $X(x,p)$ using (\ref{taylor}) we will have
\begin{eqnarray} \nonumber
&& X=\sum^{\infty}_{n=0}(-1)^n \exp\big[t(n+l)(\beta\cdot p -\xi -\frac{1}{2}\varpi :\Sigma)\big]= \\
&& \sum^{\infty}_{n=0}(-1)^n \exp\big[t(n+l)(\beta\cdot p -\xi)\big]\sum^{\infty}_{m=0}\frac{1}{m!} \big(t(n+l)(-\frac{1}{2}\varpi :\Sigma)\big)^{m}
\label{Xtaylor}
\end{eqnarray}
The product $\Sigma^{\alpha\beta}\Sigma^{\gamma\delta}$ can be decomposed to the basis of the space of $4\times 4$ matrices
\begin{eqnarray} \nonumber
&& \Sigma^{\alpha\beta}\Sigma^{\gamma\delta}=\frac{1}{4}(g^{\alpha\gamma}g^{\beta\delta}-g^{\alpha\delta}g^{\beta\gamma})I+\frac{i}{4}\epsilon^{\alpha\beta\gamma\delta}\gamma^{5}- \\
&& \frac{i}{2}(g^{\beta\delta}\Sigma^{\alpha\gamma}+g^{\alpha\gamma}\Sigma^{\beta\delta}-g^{\alpha\delta}\Sigma^{\beta\gamma}-g^{\beta\gamma}\Sigma^{\alpha\delta})\,.
\label{SigmaSigma}
\end{eqnarray}
From (\ref{SigmaSigma}) the product $(\varpi:\Sigma)^2$ can be defined. After extraction of chiral projective operators one obtains
\begin{eqnarray} \nonumber
&& (\varpi:\Sigma)^2=\eta\frac{1+\gamma^5}{2}+\theta\frac{1-\gamma^5}{2}\,,\\
&& \eta=\frac{1}{2}\varpi:\varpi +\frac{i}{2}\varpi:\widetilde{\varpi}\,,
\quad \theta=\eta^*=\frac{1}{2}\varpi:\varpi -\frac{i}{2}\varpi:\widetilde{\varpi}\,.
\label{wSigma2}
\end{eqnarray}
Then the even power of $(\varpi:\Sigma)$ can be calculated
\begin{eqnarray}
(\varpi:\Sigma)^{2k}=\eta^k\frac{1+\gamma^5}{2}+\theta^k\frac{1-\gamma^5}{2}\,,\quad k=0,1,2...\,.
\label{wSigma2k}
\end{eqnarray}
From (\ref{wSigma2k}) and (\ref{SigmaSigma}) one has
\begin{eqnarray} \nonumber 
&& \mathrm{tr}\big((\varpi:\Sigma)^{2k+1}\Sigma^{\nu\beta}\big)=(\varpi^{\nu\beta}+i\widetilde{\varpi}^{\nu\beta})\eta^k + (\varpi^{\nu\beta}-i\widetilde{\varpi}^{\nu\beta})\theta^k\,, \\
&& \mathrm{tr}\big((\varpi:\Sigma)^{2k}\Sigma^{\nu\beta}\big)=0\,,\quad k=0,1,2...\,.
\label{trwSigma2k1}
\end{eqnarray}
The traces in (\ref{axial mean}) now can be defined using decomposition (\ref{Xtaylor}) and (\ref{trwSigma2k1})
\begin{eqnarray} \nonumber 
&& \mathrm{tr}\big(X\Sigma^{\nu\beta}\big)= \sum^{\infty}_{n=0}(-1)^n \exp\big[t(n+l)(\beta\cdot p -\xi)\big]\sum^{\infty}_{k=0}\frac{1}{(2k+1)!} \big[(-\frac{1}{2})t(n+l)\big]^{2k+1}\\ 
&& \big\{(\varpi^{\nu\beta}
+i\widetilde{\varpi}^{\nu\beta})\eta^k + (\varpi^{\nu\beta}-i\widetilde{\varpi}^{\nu\beta})\theta^k\big\}\,.
\label{trXSigma1}
\end{eqnarray}
Square roots of $\eta$ and $\theta$ are
\begin{eqnarray} \nonumber 
&& \sqrt{\eta}=2g_1+2 i\, \mathrm{sgn}(\varpi:\widetilde{\varpi}) g_2\,,\quad\sqrt{\theta}=2g_1-
2 i\, \mathrm{sgn}(\varpi:\widetilde{\varpi}) g_2\,, \\ \nonumber
&& g_1=\frac{1}{4}\big(\sqrt{(\varpi:\varpi)^2+(\varpi:\widetilde{\varpi})^2}+\varpi:\varpi\big)^{1/2}\,, \\
&& g_2=\frac{1}{4}\big(\sqrt{(\varpi:\varpi)^2+(\varpi:\widetilde{\varpi})^2}-\varpi:\varpi\big)^{1/2}\,.
\label{sqrt}
\end{eqnarray}
Now, the series over $k$ in (\ref{trXSigma1}) can be summed to sine functions
\begin{eqnarray} \nonumber 
&& \mathrm{tr}\big(X\Sigma^{\nu\beta}\big)=\sum^{\infty}_{n=0}(-1)^n \exp\big[t(n+l)(\beta\cdot p -\xi)\big] \\ 
&& i \Big\{\frac{1}{\sqrt{\eta}}\sin\big(\frac{i}{2}t(n+l)\sqrt{\eta}\big)[\varpi^{\nu\beta}
+i\widetilde{\varpi}^{\nu\beta}] + \frac{1}{\sqrt{\theta}}\sin\big(\frac{i}{2}t(n+l)\sqrt{\theta}\big)[\varpi^{\nu\beta}-i\widetilde{\varpi}^{\nu\beta}]\Big\}\,.
\label{trXSigma2}
\end{eqnarray}
Sines in (\ref{trXSigma2}) can be decomposed to exponents and after that the series over $n$ can be summed back using (\ref{taylor}). Also we notice, that $\mathrm{sgn}(\varpi:\widetilde{\varpi})$ can be extracted as a factor of $\widetilde{\varpi}$
\begin{eqnarray} \nonumber 
&& \mathrm{tr}\big(X\Sigma^{\nu\beta}\big)=\Big\{\big(\exp\big[(\beta\cdot p -\xi -g_1+i g_2)\big]+1\big)^{-1} 
-\big(\exp\big[(\beta\cdot p -\xi +g_1-i g_2)\big]+1\big)^{-1}\Big\} \\ \nonumber 
&& \frac{1}{4(g_1-i g_2)}[\varpi^{\nu\beta}
-i\,\mathrm{sgn}(\varpi:\widetilde{\varpi})\widetilde{\varpi}^{\nu\beta}]+\\ \nonumber 
&& \Big\{\big(\exp\big[(\beta\cdot p -\xi - g_1-i g_2)\big]+1\big)^{-1} -
\big(\exp\big[(\beta\cdot p -\xi +g_1+i g_2)\big]+1\big)^{-1}\Big\}\\
&&\frac{1}{4(g_1+i g_2)}[\varpi^{\nu\beta}
+i\,\mathrm{sgn}(\varpi:\widetilde{\varpi}) \widetilde{\varpi}^{\nu\beta}]\,.
\label{trXSigma3}
\end{eqnarray}
The trace $\mathrm{tr}\big(\bar{X}\Sigma^{\nu\beta}\big)$ can be formally obtained from (\ref{trXSigma3}) by replacement $\xi\to -\xi$ and change of overall sign.

\end{document}